\documentclass[twocolumn]{article}

\usepackage{sigmod}
\usepackage{usenix}
\usepackage{authblk}
\usepackage{times}
\usepackage{macros}
\usepackage{alltt}
\usepackage{xcolor}
\usepackage{comment}
\usepackage{fancyhdr}
\usepackage{hyperref}
\usepackage{url}
\usepackage{bm}
\usepackage{graphicx}

\newcommand\Receive{\textbf{Receive} }

\title{TensorFlow:\\Large-Scale Machine Learning on Heterogeneous Distributed Systems\\[.1em]
\small{(Preliminary White Paper, November 9, 2015)}}

\author{
\large Mart\'{\i}n~Abadi,
Ashish~Agarwal,
Paul~Barham,
Eugene~Brevdo,
Zhifeng~Chen,
Craig~Citro,\newline
Greg~S.~Corrado,
Andy~Davis,
Jeffrey~Dean,
Matthieu~Devin,
Sanjay~Ghemawat,
Ian~Goodfellow,\newline
Andrew~Harp,
Geoffrey~Irving,
Michael~Isard,
Yangqing Jia,
Rafal~Jozefowicz,
Lukasz~Kaiser,\newline
Manjunath~Kudlur,
Josh~Levenberg,
Dan~Man\'{e},
Rajat~Monga,
Sherry~Moore,
Derek~Murray,\newline
Chris~Olah,
Mike~Schuster,
Jonathon~Shlens,
Benoit~Steiner,
Ilya~Sutskever,
Kunal~Talwar,\newline
Paul~Tucker,
Vincent~Vanhoucke,
Vijay~Vasudevan,
Fernanda~Vi\'{e}gas,
Oriol~Vinyals,\newline
Pete~Warden,
Martin~Wattenberg,
Martin~Wicke,
Yuan~Yu, and
Xiaoqiang~Zheng\newline
Google Research\thanks{Corresponding authors: Jeffrey Dean and Rajat
  Monga:\\{\tt \{jeff,rajatmonga\}@google.com}}
}

\date{\today}

\begin{document}

\maketitle

\begin{abstract}

\begin{small}
TensorFlow~\cite{tensorflow2015-whitepaper} is an interface for
expressing machine learning algorithms, and an implementation for
executing such algorithms.  A computation expressed using TensorFlow
can be executed with little or no change on a wide variety of
heterogeneous systems, ranging from mobile devices such as phones and
tablets up to large-scale distributed systems of hundreds of machines
and thousands of computational devices such as GPU cards.  The system
is flexible and can be used to express a wide variety of algorithms,
including training and inference algorithms for deep neural network
models, and it has been used for conducting research and for deploying
machine learning systems into production across more than a
dozen areas of computer science and other fields, including speech
recognition, computer vision, robotics, information retrieval, natural
language processing, geographic information extraction, and
computational drug discovery.  This paper describes the TensorFlow
interface and an implementation of that interface that we have built
at Google.  The TensorFlow API and a reference implementation were
released as an open-source package under the Apache 2.0 license in
November, 2015 and are available at
\href{www.tensorflow.org}{www.tensorflow.org}.
\end{small}

\end{abstract}

\section{Introduction}
The Google Brain project started in 2011 to explore the use of
very-large-scale deep neural networks, both for research and for use
in Google's products.  As part of the early work in this project, we
built DistBelief, our first-generation scalable distributed training
and inference system~\cite{Dean-et-al-NIPS2012}, and this system has
served us well.  We and others at Google have performed a wide variety
of research using DistBelief including work on unsupervised learning
\cite{QuocLe-ICML2012}, language representation \cite{Mikolov-et-al-ICLR2013, Vinyals-et-al-Grammar-2014}, models for image classification and object
detection \cite{frome2013devise, Szegedy-et-al-CVPR2015}, video
classification~\cite{karpathy2014large}, speech
recognition~\cite{Zeiler-et-al-ICASSP-2013,deepSpeechReviewSPM2012,
  heigold2013multilingual}, sequence
prediction~\cite{Sutskever-et-al-NIPS2014}, move selection for
Go~\cite{maddison2014move}, pedestrian
detection~\cite{angelova2015pedestrian}, reinforcement
learning~\cite{nair2015massively}, and other areas~\cite{gonzalez2015frame,ba2014multiple}.   In addition, often in close collaboration with the Google
Brain team, more than 50 teams at Google and other Alphabet companies
have deployed deep neural networks using DistBelief in a wide variety
of products, including Google Search~\cite{bloomberg-rankbrain2015},
our advertising products, our speech recognition
systems~\cite{googleresearch-blog-speechreognition2012,googleresearch-blog-neuralnetworksvoice2015,googleresearch-blog-voicesearch2015},
Google Photos~\cite{googleresearch-blog-improvingphotosearch2013},
Google Maps and StreetView~\cite{Goodfellow+et+al-ICLR2014a}, Google
Translate~\cite{googleresearch-blog-translateapp2015}, YouTube, and
many others.

Based on our experience with DistBelief and a more complete
understanding of the desirable system properties and requirements for
training and using neural networks, we have built TensorFlow, our
second-generation system for the implementation and deployment of
large-scale machine learning models.  TensorFlow takes computations
described using a dataflow-like model and maps them onto a wide
variety of different hardware platforms, ranging from running
inference on mobile device platforms such as Android and iOS to
modest-sized training and inference systems using single machines
containing one or many GPU cards to large-scale training systems
running on hundreds of specialized machines with thousands of
GPUs.  Having a single system that can span such a broad range of platforms
significantly simplifies the real-world use of machine learning system, as we have
found that having separate systems for large-scale training and small-scale
deployment leads to significant maintenance burdens and leaky
abstractions.  TensorFlow computations are expressed as stateful
dataflow graphs (described in more detail in Section~\ref{sec:programmingmodel}), and we have
focused on making the system both flexible enough for quickly
experimenting with new models for research purposes and sufficiently
high performance and robust for production training and deployment of
machine learning models.  For scaling neural network training to
larger deployments, TensorFlow allows clients to easily express
various kinds of parallelism through replication and parallel
execution of a core model dataflow graph, with many different
computational devices all collaborating to update a set of shared
parameters or other state.  Modest changes in the description of the
computation allow a wide variety of different approaches to
parallelism to be achieved and tried with low effort
\cite{Dean-et-al-NIPS2012,krizhevsky2014one,recht2011hogwild}.  Some
TensorFlow uses allow some flexibility in terms of the consistency of
parameter updates, and we can easily express and take advantage of
these relaxed synchronization requirements in some of our larger
deployments.  Compared to DistBelief, TensorFlow's programming model
is more flexible, its performance is significantly better, and it
supports training and using a broader range of models on a wider
variety of heterogeneous hardware platforms.

\begin{figure*}[t]
\label{figure:example}
\begin{small}
\begin{verbatim}
import tensorflow as tf

b = tf.Variable(tf.zeros([100]))                   # 100-d vector, init to zeroes
W = tf.Variable(tf.random_uniform([784,100],-1,1)) # 784x100 matrix w/rnd vals
x = tf.placeholder(name="x")                       # Placeholder for input
relu = tf.nn.relu(tf.matmul(W, x) + b)             # Relu(Wx+b)
C = [...]                                          # Cost computed as a function
                                                   # of Relu

s = tf.Session()
for step in xrange(0, 10):
  input = ...construct 100-D input array ...       # Create 100-d vector for input
  result = s.run(C, feed_dict={x: input})          # Fetch cost, feeding x=input
  print step, result
\end{verbatim}
\end{small}
\caption{Example TensorFlow code fragment}
\end{figure*}

\begin{figure*}
\centerline{\includegraphics[width=3.5cm]{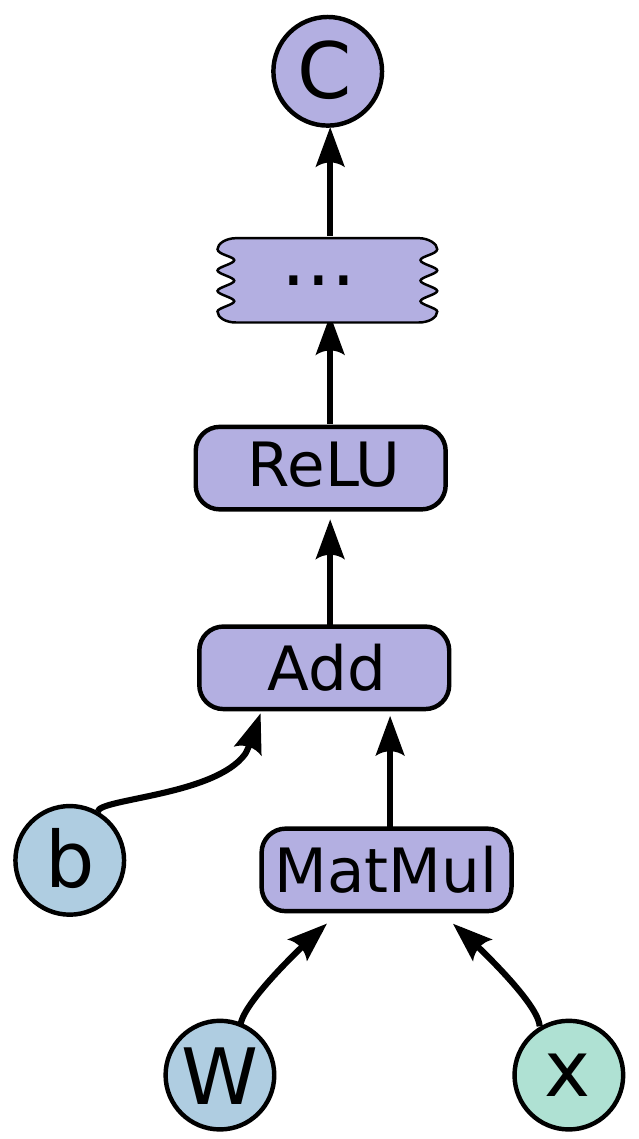}}
\caption{Corresponding computation graph for Figure~\ref{figure:example}}
\label{fig:example-graph}
\end{figure*}

Dozens of our internal clients of DistBelief have already switched to
TensorFlow. These clients rely on TensorFlow for research and production, with tasks as diverse as
running inference for computer vision models on mobile phones to
large-scale training of deep neural networks with hundreds of billions
of parameters on hundreds of billions of example records using many
hundreds of machines~\cite{bloomberg-rankbrain2015,Sutskever-et-al-NIPS2014,Szegedy-et-al-CVPR2015,googleresearch-blog-translateapp2015,vinyals2015pointer,ramsundar2015massively}.  Although these applications have concentrated
on machine learning and deep neural networks in particular, we expect
that TensorFlow's abstractions will be useful in a variety of other
domains, including other kinds of machine learning algorithms, and possibly
other kinds of numerical computations.  We have open-sourced
the TensorFlow API and a reference implementation under the
Apache 2.0 license in November, 2015, available at \href{www.tensorflow.org}{www.tensorflow.org}.

The rest of this paper describes TensorFlow in more detail.
Section~\ref{sec:programmingmodel} describes the programming
model and basic concepts of the TensorFlow interface, and
Section~\ref{sec:implementation} describes both our
single machine and distributed implementations.
Section~\ref{sec:extensions} describes several extensions to the basic
programming model, and Section~\ref{sec:optimizations} describes
several optimizations to the basic implementations.
Section~\ref{sec:experience} describes some of our experiences in using
TensorFlow, Section~\ref{sec:idioms} describes several programming
idioms we have found helpful when using TensorFlow, and Section~\ref{sec:tools} describes several auxiliary tools we have
built around the core TensorFlow system.
Sections~\ref{sec:futurework} and~\ref{sec:relatedwork} discuss future
and related work, respectively, and Section~\ref{sec:conclusions}
offers concluding thoughts.

\section{Programming Model and Basic Concepts}
\label{sec:programmingmodel}

A TensorFlow computation is described by a directed \emph{graph}, which is
composed of a set of \emph{nodes}.  The graph represents a dataflow
computation, with extensions for allowing some kinds of nodes to
maintain and update persistent state and for branching and looping
control structures within the graph in a manner similar to Naiad
\cite{murray2013naiad}.  Clients typically construct a computational
graph using one of the supported frontend languages (C++ or Python).
An example fragment to construct and then execute a TensorFlow graph
using the Python front end is shown in Figure~\ref{figure:example},
and the resulting computation graph in Figure~\ref{fig:example-graph}.

In a TensorFlow graph, each \emph{node} has zero or more inputs and
zero or more outputs, and represents the instantiation of an
\emph{operation}.  Values that flow along normal edges in the graph
(from outputs to inputs) are \emph{tensors}, arbitrary dimensionality
arrays where the underlying element type is specified or inferred at
graph-construction time.  Special edges, called \emph{control
  dependencies}, can also exist in the graph: no data flows along such
edges, but they indicate that the source node for the control
dependence must finish executing before the destination node for the
control dependence starts executing.  Since our model includes mutable
state, control dependencies can be used directly by clients to enforce
“happens before” relationships.  Our implementation also sometimes
inserts control dependencies to enforce orderings between otherwise
independent operations as a way of, for example, controlling the peak
memory usage.

\subsubsection*{Operations and Kernels}

An \emph{operation} has a name and represents an abstract computation
(e.g., ``matrix multiply'', or ``add'').  An operation can have \emph{attributes},
and all attributes must be provided or inferred at graph-construction
time in order to instantiate a node to perform the operation.  One
common use of attributes is to make operations polymorphic over
different tensor element types (e.g., “add of two tensors of type
float” versus “add of two tensors of type int32”).  A \emph{kernel} is a
particular implementation of an operation that can be run on a
particular type of device (e.g., CPU or GPU).  A TensorFlow binary
defines the sets of operations and kernels available via a
registration mechanism, and this set can be extended by linking in
additional operation and/or kernel definitions/registrations.  Table
\ref{table:operations} shows some of the kinds of operations built
into the core TensorFlow library.

\begin{table*}
\centering
\small
\begin{center}
\begin{tabular}{|l|l|}
\hline
{\bf Category} &
{\bf Examples} \\
\hline
Element-wise mathematical operations &
Add, Sub, Mul, Div, Exp, Log, Greater, Less, Equal, ... \\
Array operations &
Concat, Slice, Split, Constant, Rank, Shape, Shuffle, ... \\
Matrix operations &
MatMul, MatrixInverse, MatrixDeterminant, ... \\
Stateful operations &
Variable, Assign, AssignAdd, ... \\
Neural-net building blocks &
SoftMax, Sigmoid, ReLU, Convolution2D, MaxPool, ... \\
Checkpointing operations &
Save, Restore \\
Queue and synchronization operations &
Enqueue, Dequeue, MutexAcquire, MutexRelease, ... \\
Control flow operations &
Merge, Switch, Enter, Leave, NextIteration \\
\hline
\end{tabular}
\end{center}
\caption{\small Example TensorFlow operation types}
\label{table:operations}
\end{table*}

\subsubsection*{Sessions}

Clients programs interact with the TensorFlow system by creating a
\emph{Session}.  To create a computation graph, the Session interface
supports an \emph{Extend} method to augment the current graph managed
by the session with additional nodes and edges (the initial graph when
a session is created is empty).  The other primary operation supported
by the session interface is \emph{Run}, which takes a set of output
names that need to be computed, as well as an optional set of tensors
to be “fed” into the graph in place of certain outputs of nodes.
Using the arguments to Run, the TensorFlow implementation can compute
the transitive closure of all nodes that must be executed in order to
compute the outputs that were requested, and can then arrange to
execute the appropriate nodes in an order that respects their
dependencies (as described in more detail in
\ref{sec:basic-execution}).  Most of our uses of TensorFlow set up a
Session with a graph once, and then execute the full graph or a few
distinct subgraphs thousands or millions of times via Run calls.

\subsubsection*{Variables}

In most computations a graph is executed multiple times.  Most tensors
do not survive past a single execution of the graph.  However, a
\emph{Variable} is a special kind of operation that returns a handle
to a persistent mutable tensor that survives across executions of a graph.
Handles to these persistent mutable tensors can be passed to a handful
of special operations, such as \emph{Assign} and \emph{AssignAdd}
(equivalent to ‘+=’) that mutate the referenced tensor.
For machine learning applications of TensorFlow, the
parameters of the model are typically stored in tensors held in
variables, and are updated as part of the \emph{Run} of the training graph
for the model.

\section{Implementation}
\label{sec:implementation}

\begin{figure*}
\includegraphics[width=15cm]{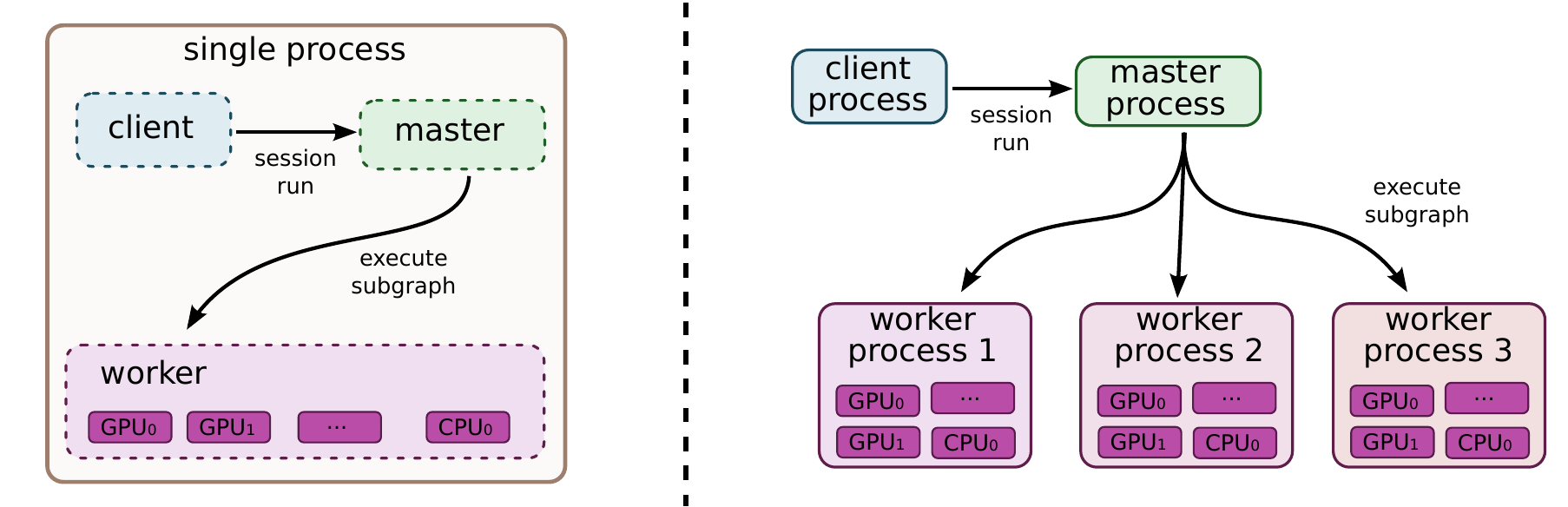}
\caption{Single machine and distributed system structure}
\label{fig:system-diagram}
\end{figure*}

The main components in a TensorFlow system are the \emph{client}, which uses
the Session interface to communicate with the \emph{master}, and one or more
\emph{worker processes}, with each worker process responsible for
arbitrating access to one or
more computational \emph{devices} (such as CPU cores or GPU cards) and for
executing graph nodes on those devices as instructed by the
master.  We have both \emph{local} and \emph{distributed} implementations of the
TensorFlow interface.  The local implementation is used when the
client, the master, and the worker all run on a single machine in the
context of a single operating system process (possibly with multiple
devices, if for example, the machine has many GPU cards
installed).  The distributed implementation shares most of the code
with the local implementation, but extends it with support for an
environment where the client, the master, and the workers can all be
in different processes on different machines.  In our distributed
environment, these different tasks are containers in jobs managed by a
cluster scheduling system \cite{verma2015large}. These two different modes are
illustrated in Figure~\ref{fig:system-diagram}.  Most of the rest of this section
discusses issues that are common to both implementations, while
Section~\ref{sec:distributed-implementation} discusses some issues that are particular to the
distributed implementation.

\subsubsection*{Devices}
Devices are the computational heart of TensorFlow.  Each worker is
responsible for one or more devices, and each device has a device
type, and a name.  Device names are composed of pieces that identify
the device's type, the device's index within the worker, and, in our
distributed setting, an identification of the job and task of the
worker (or “localhost” for the case where the devices are local to the
process).  Example device names are
\texttt{"/job:localhost/device:cpu:0"} or
\texttt{"/job:worker/task:17/device:gpu:3"}.  We have implementations
of our Device interface for CPUs and GPUs, and new device
implementations for other device types can be provided via a
registration mechanism.  Each device object is responsible for
managing allocation and deallocation of device memory, and for
arranging for the execution of any kernels that are requested by
higher levels in the TensorFlow implementation.

\subsubsection*{Tensors}

A tensor in our implementation is a typed, multi-dimensional
array.  We support a variety of tensor element types, including signed
and unsigned integers ranging in size from 8 bits to 64 bits, IEEE
float and double types, a complex number type, and a string type (an
arbitrary byte array).  Backing store of the appropriate size is
managed by an allocator that is specific to the device on which the tensor
resides.  Tensor backing store buffers are reference counted and are
deallocated when no references remain.

\subsection{Single-Device Execution}
\label{sec:basic-execution}

Let's first consider the simplest execution scenario: a single worker
process with a single device.  The nodes of the graph are executed in
an order that respects the dependencies between nodes.  In particular,
we keep track of a count per node of the number of dependencies of
that node that have not yet been executed.  Once this count drops to
zero, the node is eligible for execution and is added to a ready
queue.  The ready queue is processed in some unspecified order,
delegating execution of the kernel for a node to the device object.  When
a node has finished executing, the counts of all nodes that depend
on the completed node are decremented.

\subsection{Multi-Device Execution}

Once a system has multiple devices, there are two main complications:
deciding which device to place the computation for each node in the
graph, and then managing the required communication of data across
device boundaries implied by these placement decisions.  This
subsection discusses these two issues.

\subsubsection{Node Placement}
\label{sec:placement}

Given a computation graph, one of the main responsibilities of the
TensorFlow implementation is to map the computation onto the set of
available devices.  A simplified version of this algorithm is
presented here.  See Section~\ref{sec:constraints} for extensions
supported by this algorithm.

One input to the placement algorithm is a cost model, which contains
estimates of the sizes (in bytes) of the input and output tensors for
each graph node, along with estimates of the computation time required
for each node when presented with its input tensors.  This cost model
is either statically estimated based on heuristics associated with
different operation types, or is measured based on an actual set of
placement decisions for earlier executions of the graph.

The placement algorithm first runs a simulated execution of the graph.
The simulation is described below and ends up picking a device
for each node in the graph using greedy heuristics.  The node to
device placement generated by this simulation is also used as the
placement for the real execution.

The placement algorithm starts with the sources of the
computation graph, and simulates the activity on each device in the
system as it progresses.  For each node that is reached in this
traversal, the set of feasible devices is considered (a device may not
be feasible if the device does not provide a kernel that implements the
particular operation). For
nodes with multiple feasible devices, the placement algorithm uses a
greedy heuristic that examines the effects on the completion time of
the node of placing the node on each possible device.  This heuristic
takes into account the estimated or measured execution time of the
operation on that kind of device from the cost model, and also
includes the costs of any communication that would be introduced in
order to transmit inputs to this node from other devices to the
considered device.  The device where the node's operation would finish
the soonest is selected as the device for that operation, and the
placement process then continues onwards to make placement decisions
for other nodes in the graph, including downstream nodes that are now
ready for their own simulated execution.  Section~\ref{sec:constraints} describes
some extensions that allow users to provide hints and partial
constraints to guide the placement algorithm.  The placement algorithm is
an area of ongoing development within the system.

\subsubsection{Cross-Device Communication}
\label{sec:crossdevice}

\begin{figure}[here]
\includegraphics[width=8cm]{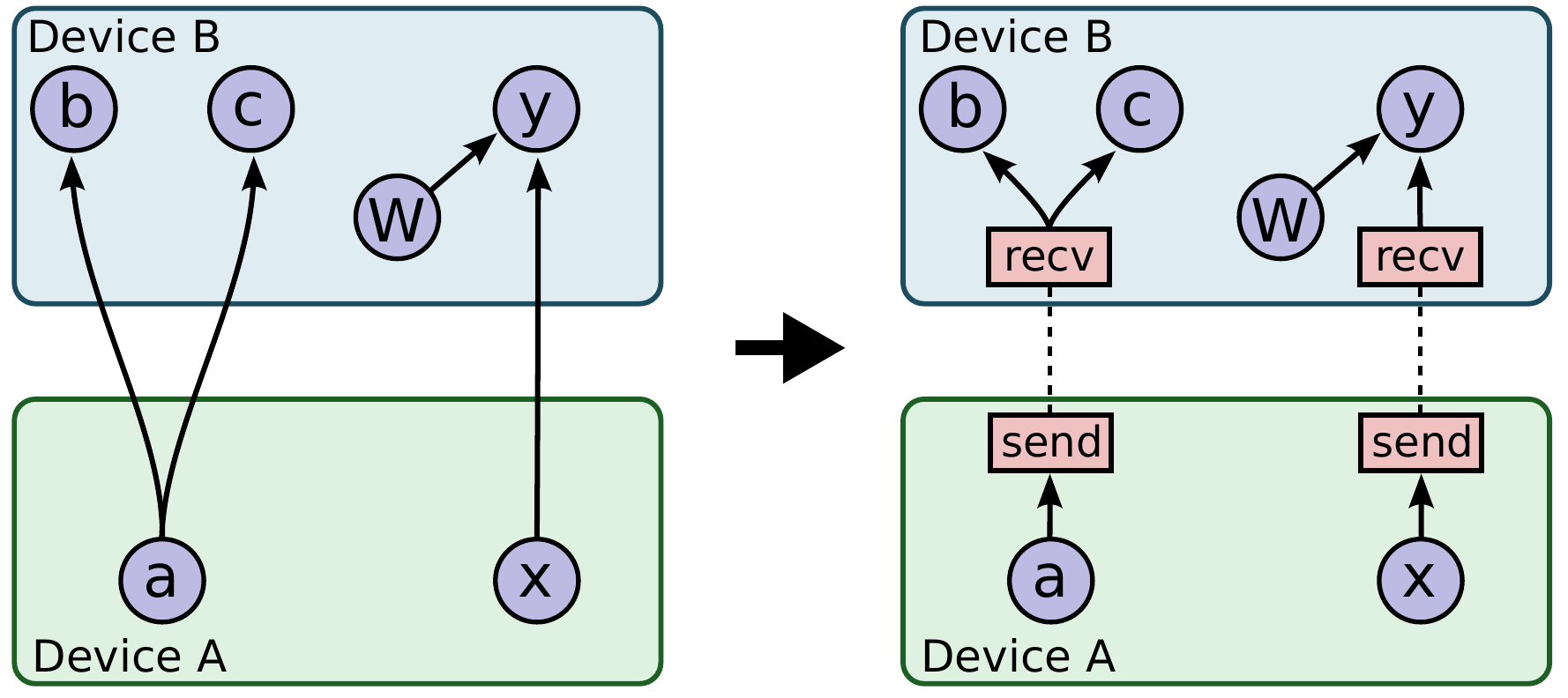}
\caption{Before \& after insertion of Send/Receive nodes}
\label{fig:sendrecv}
\end{figure}

Once the node placement has been computed, the graph is partitioned
into a set of subgraphs, one per device.  Any cross-device edge from
\textbf{x} to \textbf{y} is removed and replaced by an edge from \textbf{x} to a new \emph{Send} node
in \textbf{x}'s subgraph and an edge from a corresponding \emph{Receive} node to
\textbf{y} in \textbf{y}'s subgraph.  See Figure~\ref{fig:sendrecv} for an example of
this graph transformation.

At runtime, the implementations of the Send and Receive nodes
coordinate to transfer data across devices.  This allows us to isolate
all communication inside Send and Receive
implementations, which simplifies the rest of the runtime.

When we insert Send and Receive nodes, we canonicalize all users of a
particular tensor on a particular device to use a single Receive node,
rather than one Receive node per downstream user on a particular
device.  This ensures that the data for the needed tensor is only
transmitted once between a source device $\rightarrow$ destination device
pair, and that memory for the tensor on the destination device is only
allocated once, rather than multiple times (e.g., see nodes \textbf{b}
and \textbf{c} in Figure~\ref{fig:sendrecv})

By handling communication in this manner, we also allow the scheduling
of individual nodes of the graph on different devices to be
decentralized into the workers: the Send and Receive nodes impart
the necessary synchronization between different workers and devices,
and the master only needs to issue a single Run request per graph
execution to each worker that has any nodes for the graph, rather than
being involved in the scheduling of every node or every cross-device
communication.  This makes the system much more scalable and allows
much finer-granularity node executions than if the scheduling were
forced to be done by the master.

\subsection{Distributed Execution}
\label{sec:distributed-implementation}

Distributed execution of a graph is very similar to multi-device
execution.  After device placement, a subgraph is created per
device.  Send/Receive node pairs that communicate across worker
processes use remote communication mechanisms such as TCP or RDMA
to move data across machine boundaries.

\subsubsection*{Fault Tolerance}

Failures in a distributed execution can be detected in a variety
of places.  The main ones we rely on are (a) an error in a communication
between a Send and Receive node pair, and (b) periodic health-checks from
the master process to every worker process.

When a failure is detected, the entire graph execution is aborted
and restarted from scratch.  Recall however that Variable nodes
refer to tensors that persist across executions of the graph.  We
support consistent checkpointing and recovery of this state on
a restart.  In partcular, each Variable node is connected to a
Save node. These Save nodes are executed periodically, say once every
N iterations, or once every N seconds.
When they execute, the contents of the variables are
written to persistent storage, e.g., a distributed file system.
Similarly each Variable is connected
to a Restore node that is only enabled in the first iteration after
a restart.  See Section~\ref{sec:partialexecution} for details on
how some nodes can only be enabled on some executions of the graph.

%

\section{Extensions}
\label{sec:extensions}

In this section we describe several more advanced features of the
basic programming model that was introduced in
Section~\ref{sec:programmingmodel}.

\subsection{Gradient Computation}
\label{sec:gradientcomputation}

Many optimization algorithms, including common machine learning
training algorithms like stochastic gradient descent~\cite{rumelhart1988learning},
compute the gradient of a cost function with respect to a set of
inputs. Because this is such a common need, TensorFlow has built-in
support for automatic gradient computation. If a tensor $C$ in a
TensorFlow graph depends, perhaps through a complex subgraph of
operations, on some set of tensors $\{X_k\}$, then there is a built-in
function that will return the tensors $\{dC/dX_k\}$. Gradient tensors
are computed, like other tensors, by extending the TensorFlow graph,
using the following procedure.

\begin{figure}
\centerline{\includegraphics[width=6.5cm]{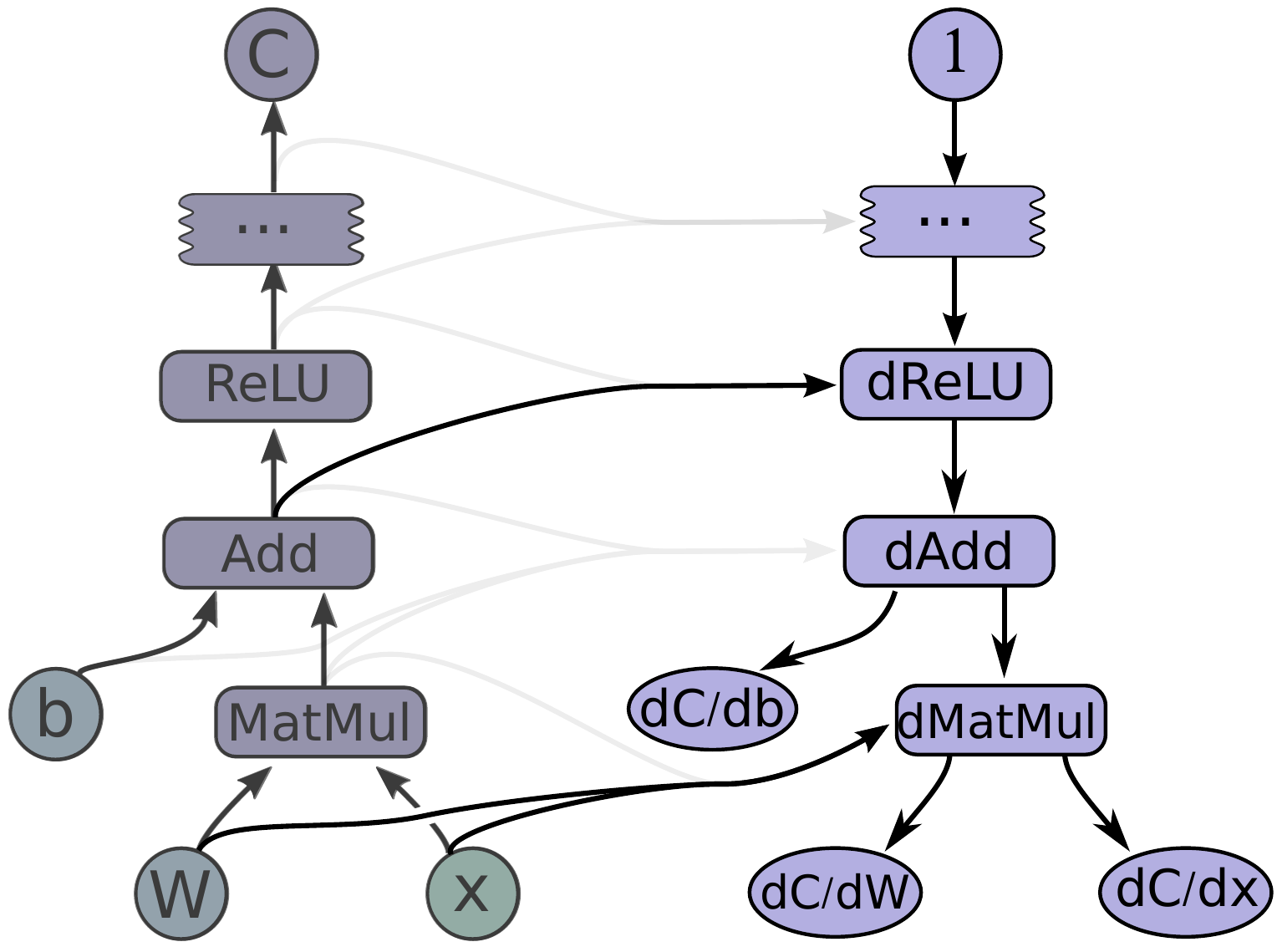}}
\caption{Gradients computed for graph in
  Figure~\ref{fig:example-graph}}
\label{fig:gradients}
\end{figure}

When TensorFlow needs to compute the gradient of a tensor $C$ with
respect to some tensor $I$ on which $C$ depends, it first finds the
path in the computation graph from $I$ to $C$. Then it backtracks from
$C$ to $I$, and for each operation on the backward path it adds a node
to the TensorFlow graph, composing the partial gradients along the
backwards path using the chain rule. The newly added node computes the
``gradient function'' for the corresponding operation in the forward
path. A gradient function may be registered by any operation.  This
function takes as input not only the partial gradients computed
already along the backward path, but also, optionally, the inputs and
outputs of the forward operation. Figure~\ref{fig:gradients} shows
gradients for a cost computed from the example of
Figure~\ref{fig:example-graph}. Grey arrows show potential inputs to
gradient functions that are not used for the particular operations
shown. The addition needed to Figure~\ref{figure:example} to compute
these gradients is:
\begin{small}
\begin{verbatim}
[db,dW,dx] = tf.gradients(C, [b,W,x])
\end{verbatim}
\end{small}
In general an operation may have multiple outputs, and $C$ may
only depend on some of them. If, for example, operation $O$ has two
outputs $y_1$ and $y_2$, and $C$ only depends on $y_2$, then the first
input to $O$'s gradient function is set to $0$ since $dC/dy_1=0$.

Automatic gradient computation complicates optimization, particularly
of memory usage. When executing ``forward'' computation subgraphs,
i.e., those that are explicitly constructed by the user, a sensible
heuristic breaks ties when deciding which node to execute next by
observing the order in which the graph was constructed. This generally
means that temporary outputs are consumed soon after being
constructed, so their memory can be reused quickly. When the heuristic
is ineffective, the user can change the order of graph construction,
or add control dependencies as described in
Section~\ref{sec:optimizations}. When gradient nodes are automatically
added to the graph, the user has less control, and the heuristics may
break down. In particular, because gradients reverse the forward
computation order, tensors that are used early in a graph's execution
are frequently needed again near the end of a gradient
computation. Such tensors can hold on to a lot of scarce GPU memory
and unnecessarily limit the size of computations. We are actively
working on improvements to memory management to deal better with such
cases. Options include using more sophisticated heuristics to
determine the order of graph execution, recomputing tensors instead of
retaining them in memory, and swapping out long-lived tensors from GPU
memory to more plentiful host CPU memory.

\subsection{Partial Execution}
\label{sec:partialexecution}

Often a client wants to execute just a subgraph of the entire
execution graph.  To support this, once the client has set up a
computation graph in a Session, our Run method allows them to execute
an arbitrary subgraph of the whole graph, and to inject arbitrary data
along any edge in the graph, and to retrieve data flowing along any
edge in the graph.

Each node in the graph has a name, and each output of a node is identified
by the source node name and the output port from the
node, numbered from 0 (e.g., ``bar:0'' refers to the 1st output of the
``bar'' node, while ``bar:1'' refers to the 2nd output).

Two arguments to the Run call help define the exact
subgraph of the computation graph that will be executed.  First, the
Run call accepts inputs, an optional mapping of
\texttt{\emph{name}:\emph{port}} names to ``fed'' tensors values.  Second, the
Run call accepts \texttt{output\_names}, a list of output
\texttt{\emph{name}[:\emph{port}]} specifications indicating which
nodes should be executed, and, if the port portion is present in a
name, that that particular output tensor value for the node should be
returned to the client if the Run call completes successfully.

\begin{figure}[here]
\includegraphics[width=8.5cm]{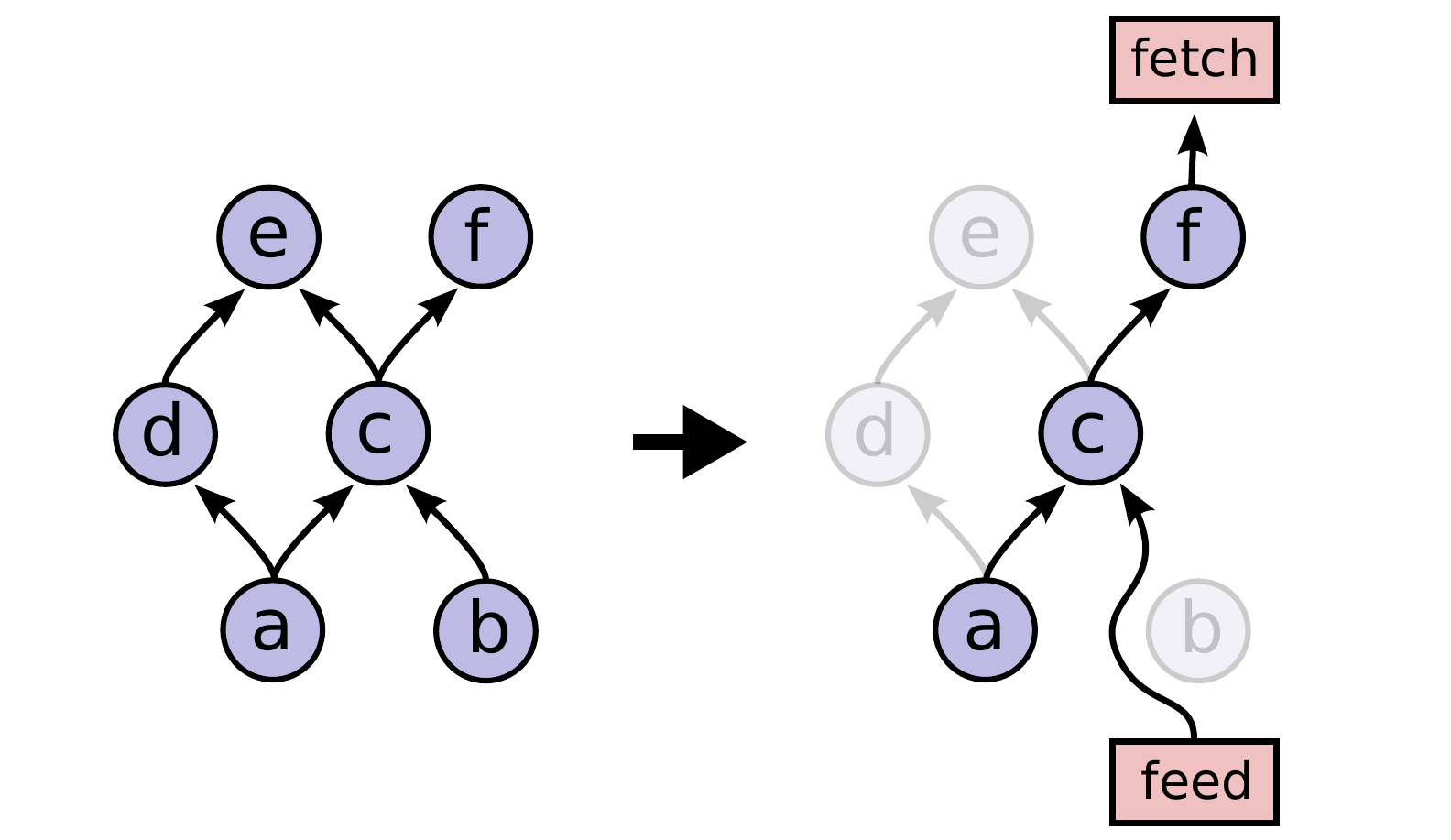}
\caption{Before and after graph transformation for partial execution}
\label{fig:feedfetch}
\end{figure}

The graph is transformed based on the values of inputs and outputs.
Each node:port specified in inputs is replaced with a \textbf{feed}
node, which will pick up the provided input tensor from
specially-initialized entries in a Rendezvous object used for the Run
call.  Similarly, each output name with a port is connected to a
special \textbf{fetch} node that arranges to save the output tensor
and return it to the client when the Run call is complete.  Finally,
once the graph has been rewritten with the insertion of these special
\textbf{feed} and \textbf{fetch} nodes, the set of nodes to execute can be
determined by starting at each of the nodes named by any output and
working backwards in the graph using the graph dependencies to
determine the full set of nodes that must be executed in the rewritten
graph in order to compute the outputs.  Figure~\ref{fig:feedfetch}
shows an original graph on the left, and the transformed graph that
results when Run is invoked with inputs==\{\textbf{b}\} and
outputs==\{\textbf{f:0}\}.  Since we only need to compute the output
of node \textbf{f}, we will not execute nodes \textbf{d} and
\textbf{e}, since they have no contribution to the output of \textbf{f}.

\subsection{Device Constraints}
\label{sec:constraints}

TensorFlow clients can control the placement of nodes on devices
by providing partial constraints for a node
about which devices it can execute on.  For example, \emph{``only place this
node on a device of type GPU''}, or \emph{``this node can be placed on
any device in \texttt{/job:worker/task:17}''}, or \emph{``Colocate
this node with the node named \texttt{variable13}''}.  Within the
confines of these constraints, the placement algorithm is responsible
for choosing an assignment of nodes to devices that provides fast
execution of the computation and also satisfies various constraints
imposed by the devices themselves, such as limiting the total amount
of memory needed on a device in order to execute its subset of graph
nodes.

Supporting such constraints requires changes to the placement
algorithm described in Section~\ref{sec:placement}.
We first compute the feasible set of devices for
each node, and then use union-find on the graph of colocation constraints
to compute the graph components that must be placed together. For each
such component, we compute the intersection of the feasible device
sets.  The computed feasible device set per node fits easily into the
placement algorithm's simulator.

\subsection{Control Flow}

Although dataflow graphs without any explicit control flow are quite expressive,
we have observed a number of cases where supporting conditionals and
loops can lead to more concise and efficient representations of machine
learning algorithms.

Much as in the dataflow-machine approach described by Arvind
\cite{arvind1986}, we introduce a small set of primitive control flow
operators into TensorFlow and generalize TensorFlow to handle cyclic
dataflow graphs. The \emph{Switch} and \emph{Merge} operators allow us
to skip the execution of an entire subgraph based on the value of a
boolean tensor. The \emph{Enter}, \emph{Leave}, and
\emph{NextIteration} operators allow us to express iteration.
High-level programming constructs such as if-conditionals and
while-loops can be easily compiled into dataflow graphs with these
control flow operators.

The TensorFlow runtime implements a notion of
tags and frames conceptually similar to the MIT Tagged-Token machine
\cite{arvind1990}.  Each iteration of a loop is uniquely identified
by a tag, and its execution state is represented by a frame. An input
can enter an iteration whenever it becomes available; thus, multiple
iterations can be executed concurrently.

TensorFlow uses a distributed coordination mechanism to execute graphs
with control flow.  In general, a loop can contain nodes that are
assigned to many different devices. Therefore, managing the state of a
loop becomes a problem of distributed termination
detection. TensorFlow's solution is based on graph rewriting. During
the graph partitioning, we automatically add control nodes to each
partition. These nodes implement a small state machine that
orchestrates the start and termination of each iteration, and decides
the termination of the loop. For each iteration, the device that owns
the loop termination predicate sends a tiny control message to every
participating device.

As explained above, we often train machine learning models by gradient
descent, and represent gradient computations as part of dataflow
graphs. When a model includes control-flow operations, we must account
for them in the corresponding gradient computation. For example, the
gradient computation for a model with an if-conditional will need to
know which branch of the conditional was taken, then apply the
gradient logic to this branch. Similarly, the gradient computation for
a model with a while-loop will need to know how many iterations were
taken, and will also rely on the intermediate values computed during
those iterations.  The basic technique is to rewrite the graph so to
memorize the values needed for the gradient computation. We omit the
somewhat intricate details of this encoding.


\subsection{Input Operations}

Although input data can be provided to a computation via feed nodes,
another common mechanism used for training large-scale machine
learning models is to have special input operation nodes in the graph,
which are typically configured with a set of filenames and which yield
a tensor containing one or more examples from the data stored in that
set of files each time they are executed.  This allows data to be read
directly from the underlying storage system into the memory of the
machine that will perform subsequent processing on the data.  In
configurations where the client process is separate from the worker
process, if the data were fed, it typically would require an extra
network hop (from the storage system to the client and then from the
client to the worker vs. directly from the storage system to ther
worker when using an input node).

\subsection{Queues}

Queues are a useful feature that we have added to TensorFlow.  They
allow different portions of the graph to execute asynchronously,
possibly at different candences, and to hand off data through Enqueue
and Dequeue operations.  Enqueue operations can block until space
becomes available in the queue, and Dequeue operations can block until
a desired minimum number of elements are available in the queue.  One
use of queues is to allow input data to be prefetched from disk files
while a previous batch of data is still being processed by the
computational portion of a machine learning model.  They can also be
used for other kinds of grouping, including accumulating many
gradients in order to compute some more complex combination of
gradients over a larger batch, or to group different input sentences
for recurrent language models into bins of sentences that are
approximately the same length, which can then be processed more
efficiently.

In addition to normal FIFO queues, we have also implemented a
shuffling queue, which randomly shuffles its elements within a large
in-memory buffer.  This shuffling functionality is useful for machine
learning algorithms that want to randomize the order in which they
process examples, for example.

%

\subsection{Containers}  

A \emph{Container} is the mechanism within TensorFlow
for managing longer-lived mutable state.  The backing store for a
\emph{Variable} lives in a container.  The default container is one
that persists until the process terminates, but we also allow
other named containers.  A container can be reset by clearing it of
its contents entirely.  Using containers, it is possible to share
state even across completely disjoint computation graphs associated
with different Sessions.

\section{Optimizations}
\label{sec:optimizations}

In this section, we describe some of the optimizations in the
TensorFlow implementation that improve performance or resource usage
of the system.

\subsection{Common Subexpression Elimination}

Since the construction of computation graphs is often done by many
different layers of abstractions in the client code, computation
graphs can easily end up with redundant copies of the same
computation.  To handle this, we have implemented a common
subexpression pass similar to the algorithm described by
Click~\cite{click1995global} that runs over the computation graph and
canonicalizes multiple copies of operations with identical inputs and
operation types to just a single one of these nodes, and redirects
graph edges appropriately to reflect this canonicalization.

%

\subsection{Controlling Data Communication and Memory Usage}

Careful scheduling of TensorFlow operations can result in better
performance of the system, in particular with respect to data
transfers and memory usage. Specifically, scheduling can reduce the
time window during which intermediate results need to be kept in
memory in between operations and hence the peak memory
consumption. This reduction is particularly important for GPU devices
where memory is scarce. Furthermore, orchestrating the communication
of data across devices can reduce contention for network resources.

While there are many opportunities for scheduling optimizations, here
we focus on one that we found particularly necessary and effective. It
concerns the scheduling of Receive nodes for reading remote values. If
no precautions are taken, these nodes may start much earlier than
necessary, possibly all at once when execution starts. By performing
an as-soon-as-possible/as-late-as-possible (ASAP/ALAP) calculation, of
the kind common in operations research, we analyze the critical paths
of graphs, in order to estimate when to start the Receive nodes. We
then insert control edges with the aim of delaying the start of these
nodes until just before their results are needed.

\subsection{Asynchronous Kernels}

In addition to normal synchronous kernels that complete their
execution at the end of the Compute method, our framework also
supports non-blocking kernels.  Such non-blocking kernels use a
slightly different interface whereby the Compute method is passed a
continuation that should be invoked when the kernel's execution is
complete.  This is an optimization for environments where having many
active threads is relatively expensive in terms of memory usage or
other resources, and allows us to avoid tying up an execution thread
for unbounded periods of time while waiting for I/O or other events to
occur.  Examples of asynchronous kernels include the \Receive kernel,
and the \textbf{Enqueue} and \textbf{Dequeue} kernels (which might need to block if
queue space is not available or if no data is available to be read,
respectively).

\subsection{Optimized Libraries for Kernel Implementations}

We often make use of pre-existing highly-optimized numerical libraries
to implement kernels for some operations.  For example, there are a
number of optimized libraries for performing matrix multiplies on
different devices, including BLAS~\cite{dongarra1990set} and
cuBLAS~\cite{nvidia2008cublas}, or GPU libraries for convolutional
kernels for deep neural nets such as
cuda-convnet~\cite{krizhevsky2014cuda} and
cuDNN~\cite{chetlur2014cudnn}.  Many of our kernel implementations are
relatively thin wrappers around such optimized libraries.

We make fairly extensive use of the open-source Eigen linear algebra
library~\cite{eigen} for many of the kernel implementations in the
system.  As one part of the development of TensorFlow, our team
(primarily Benoit Steiner) has extended the open source Eigen library
with support for arbitrary dimensionality tensor operations.

\subsection{Lossy Compression}

Some machine learning algorithms, including those typically used for
training neural networks, are tolerant of noise and reduced precision
arithmetic.  In a manner similar to the DistBelief
system~\cite{Dean-et-al-NIPS2012}, we often use lossy compression of
higher precision internal representations when sending data between
devices (sometimes within the same machine but especially across machine boundaries).
For example, we often insert special conversion nodes that convert
32-bit floating point representations into a 16-bit floating point
representation (not the proposed IEEE 16-bit floating point standard,
but rather just a 32-bit IEEE 794 float format, but with 16 bits less
precision in the mantissa), and then convert back to a 32-bit
representation on the other side of the communication channel (by just
filling in zeroes for the lost portion of the mantissa, since that's
less computationally expensive than doing the mathematically correct
probabilistic rounding when doing this $32 \rightarrow 16 \rightarrow 32$-bit conversion).

\section{Status and Experience}
\label{sec:experience}

The TensorFlow interface and a reference implementation have been open
sourced under an Apache 2.0 license, and the system is available for
download at
\href{www.tensorflow.org}{www.tensorflow.org}.  The system includes detailed
documentation, a number of tutorials, and a number of examples
demonstrating how to use the system for a variety of different machine
learning tasks.  The examples include models for classifying
hand-written digits from the MNIST dataset (the ``hello world'' of
machine learning algorithms)~\cite{lecun1998mnist}, classifying images from the
CIFAR-10 dataset~\cite{krizhevsky-cifar}, doing language modeling using a recurrent
LSTM~\cite{hochreiter1997long} network, training word embedding
vectors~\cite{Mikolov-et-al-ICLR2013} and more.

The system includes front-ends for specifying TensorFlow computations
in Python and C++, and we expect other front-ends to be added over
time in response to the desires of both internal Google users and the
broader open-source community.

We have quite a few machine learning models in our previous DistBelief
system~\cite{Dean-et-al-NIPS2012} that we have migrated over to
TensorFlow.  The rest of this section discusses some lessons we have
learned that are generalizable for any such migration of machine
learning models from one system to another, and therefore may be
valuable to others.

In particular, we focus on our lessons from porting a state-of-the-art
convolutional neural network for image recognition termed {\it
  Inception}~\cite{DBLP:journals/corr/IoffeS15}. This image
recognition system classifies $224 \times 224$ pixel images into one
of 1000 labels (e.g., ``cheetah'', ``garbage truck'', etc.). Such a
model comprises 13.6 million learnable parameters and 36,000
operations when expressed as a TensorFlow graph. Running inference on
a single image requires 2 billion multiply-add operations.

\begin{figure*}[th]
\centerline{\includegraphics[width=12cm]{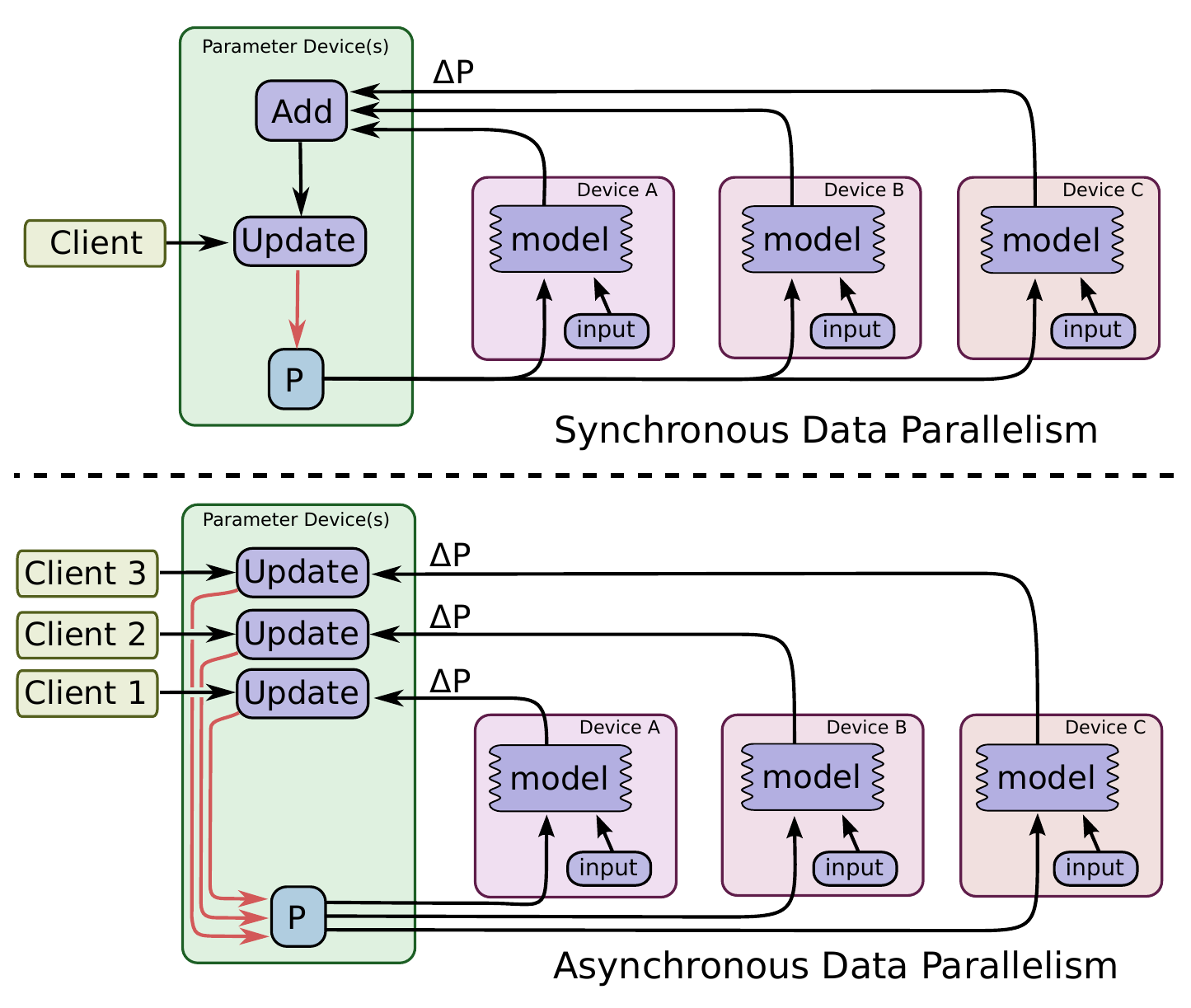}}
\caption{Synchronous and asynchronous data parallel training}
\label{fig:sync-async}
\end{figure*}

After building all necessary mathematical operations in TensorFlow,
assembling and debugging all 36,000 operations into the correct graph structure
proved challenging. Validating correctness is a difficult enterprise
because the system is inherently stochastic and only intended to behave
in a certain way in expectation --- potentially after hours of computation.
Given these circumstances, we found the following strategies critical for
porting the Inception model to TensorFlow:

\begin{enumerate}
\item {\it Build tools to gain insight into the exact number of parameters in a given model.} Such tools demonstrated subtle flaws in a complex network architecture specification. In particular we were able to identify operations and variables instantiated incorrectly due to automatic broadcasting in a mathematical operation across a dimension.
\item {\it Start small and scale up.} The first convolutional neural network that we ported from our previous system was a small network employed on the CIFAR-10 data set~\cite{krizhevsky-cifar}. Debugging such a network elucidated subtle edge cases in individual operations (e.g., max-pooling) within the machine learning system that would have been practically indecipherable in more complex models.
\item {\it Always ensure that the objective (loss function) matches between machine learning systems when learning is turned off.} Setting the learning rate to be zero helped us identify unexpected behavior in how we had randomly initialized variables in a model. Such an error would have been difficult to identify in a dynamic, training network.
\item {\it Make a single machine implementation match before debugging a distributed implementation.} This strategy helped us delineate and debug discrepancies in training performance between machine learning system. In particular, we identified bugs due to race conditions and non-atomic operations incorrectly assumed to be atomic.
\item {\it Guard against numerical errors.} Numerical libraries are inconsistent in how they handle non-finite floating point values. Convolutional neural networks are particularly susceptible to numerical instability and will tend to diverge quite regularly during experimentation and debugging phases. Guarding against this behavior by checking for non-finite floating point values allows one to detect errors in real time as opposed to identifying divergent behavior post-hoc.
\item {\it Analyze pieces of a network and understand the magnitude of
  numerical error.} Running subsections of a neural network in
  parallel on two machine learning systems provides a precise method
  to ensure that a numerical algorithm is identical across two
  systems. Given that such algorithms run with floating point
  precision, it is important to predict and understand the magnitude of
  expected numerical error in order to judge whether a given component
  is correctly implemented  (e.g., distinguishing between {\it ``within
    1e-2, great!''} and {\it ``within 1e-2: why is it so incorrect?!''}).
\end{enumerate}

Validating complex mathematical operations in the presence of an
inherently stochastic system is quite challenging. The strategies
outlined above proved invaluable in gaining confidence in the system
and ultimately in instantiating the Inception model in TensorFlow.
The end result of these efforts resulted in a 6-fold speed improvement
in training time versus our existing DistBelief implementation of the
model and such speed gains proved indispensable in training a new
class of larger-scale image recognition models.

\section{Common Programming Idioms}
\label{sec:idioms}

TensorFlow's basic dataflow graph model can be used in a variety of
ways for machine learning applications.  One domain we care about is
speeding up training of computationally intensive neural network
models on large datasets.  This section describes several techniques
that we and others have developed in order to accomplish this, and
illustrates how to use TensorFlow to realize these various approaches.

The approaches in this subsection assume that the model is being
trained using stochastic gradient descent (SGD) with relatively
modest-sized mini-batches of 100 to 1000 examples.

\subsubsection*{Data Parallel Training}

One simple technique for speeding up SGD is to parallelize the
computation of the gradient for a mini-batch across mini-batch
elements.  For example, if we are using a mini-batch size of 1000
elements, we can use 10 replicas of the model to each compute the
gradient for 100 elements, and then combine the gradients and apply
updates to the parameters synchronously, in order to behave exactly as
if we were running the sequential SGD algorithm with a batch size of
1000 elements.  In this case, the TensorFlow graph simply has many
replicas of the portion of the graph that does the bulk of the model
computation, and a single client thread drives the entire training
loop for this large graph.  This is illustrated in the top portion of
Figure~\ref{fig:sync-async}.

This approach can also be made asynchronous, where the TensorFlow
graph has many replicas of the portion of the graph that does the bulk
of the model computation, and each one of these replicas also applies
the parameter updates to the model parameters asynchronously.  In this
configuration, there is one client thread for each of the graph
replicas.  This is illustrated in the bottom portion of
Figure~\ref{fig:sync-async}.  This asynchronous approach was
also described in~\cite{Dean-et-al-NIPS2012}.

\subsubsection*{Model Parallel Training}

\begin{figure}
\centerline{\includegraphics[width=6.5cm]{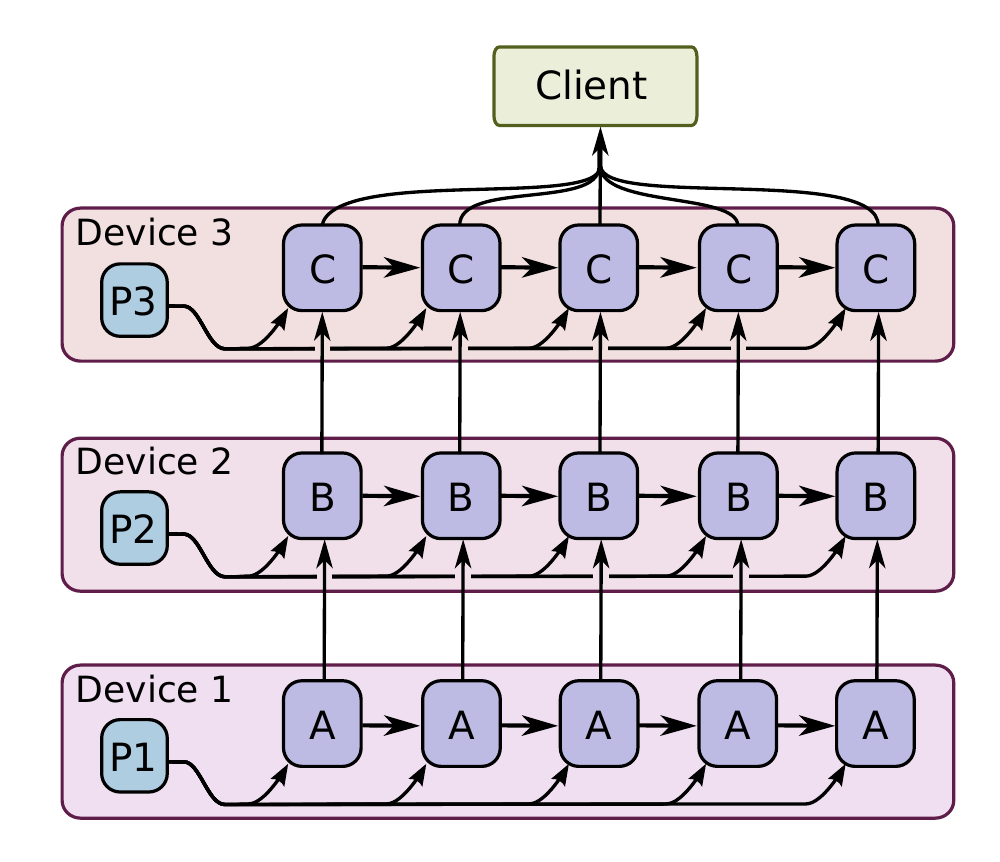}}
\caption{Model parallel training}
\label{fig:model-parallel}
\end{figure}

Model parallel training, where different portions of the model
computation are done on different computational devices simultaneously
for the same batch of examples, is also easy to express in TensorFlow.
Figure~\ref{fig:model-parallel} shows an example of a recurrent, deep
LSTM model used for sequence to sequence learning
(see~\cite{Sutskever-et-al-NIPS2014}), parallelized across three
different devices.

\subsubsection*{Concurrent Steps for Model Computation Pipelining}

\begin{figure}
\centerline{\includegraphics[width=6.5cm]{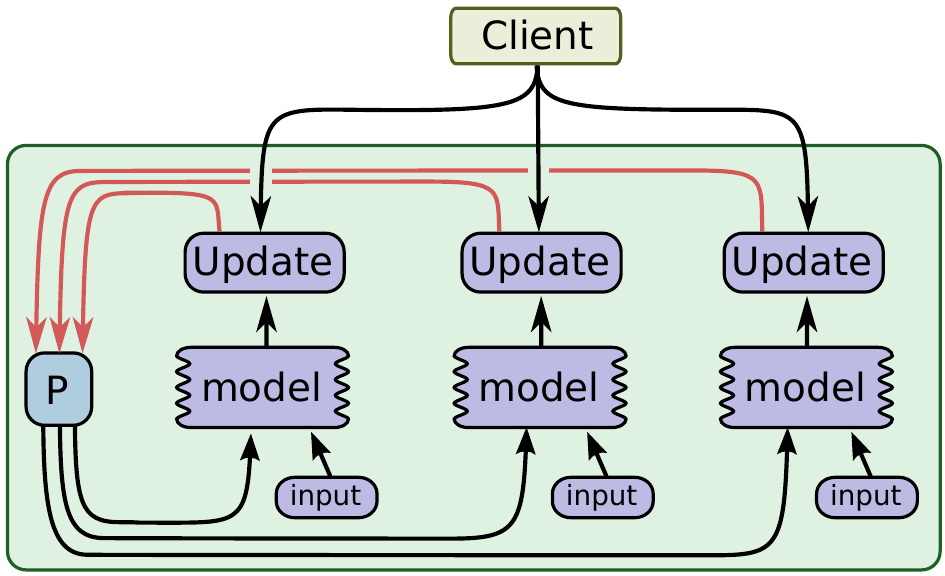}}
\caption{Concurrent steps}
\label{fig:concurrent-steps}
\end{figure}

Another common way to get better utilization for training deep neural
networks is to pipeline the computation of the model within the same
devices, by running a small number of concurrent steps within the same
set of devices.  This is shown in Figure~\ref{fig:concurrent-steps}.
It is somewhat similar to asynchronous data parallelism, except that
the parallelism occurs within the same device(s), rather than
replicating the computation graph on different devices.  This allows
``filling in the gaps'' where computation of a single batch of
examples might not be able to fully utilize the full parallelism on
all devices at all times during a single step.

\section{Performance}
\label{sec:performance}

{\it A future version of this white paper will have a comprehensive
  performance evaluation section of both the single machine and
  distributed implementations.}


%
%
%
%

\section{Tools}
\label{sec:tools}

This section describes some tools we have developed that sit alongside
the core TensorFlow graph execution engine.

\subsection{TensorBoard: Visualization of graph structures and summary
  statistics}

In order to help users understand the structure of their computation
graphs and also to understand the overall behavior of machine learning
models, we have built TensorBoard, a companion visualization tool for
TensorFlow that is included in the open source release.

\subsubsection*{Visualization of Computation Graphs}

Many of the computation graphs for deep neural networks can be quite
complex. For example, the computation graph for training a model
similar to Google's Inception model~\cite{Szegedy-et-al-CVPR2015}, a
deep convolutional neural net that had the best classification
performance in the ImageNet 2014 contest, has over 36,000 nodes in its
TensorFlow computation graph, and some deep recurrent LSTM models for
language modeling have more than 15,000 nodes.

Due to the size and topology of these graphs, naive visualization
techniques often produce cluttered and overwhelming diagrams. To help
users see the underlying organization of the graphs, the algorithms in
TensorBoard collapse nodes into high-level blocks, highlighting groups
with identical structures. The system also separates out high-degree
nodes, which often serve bookkeeping functions, into a separate area
of the screen. Doing so reduces visual clutter and focuses attention
on the core sections of the computation graph.

\begin{figure*}
\centerline{\includegraphics[width=13cm]{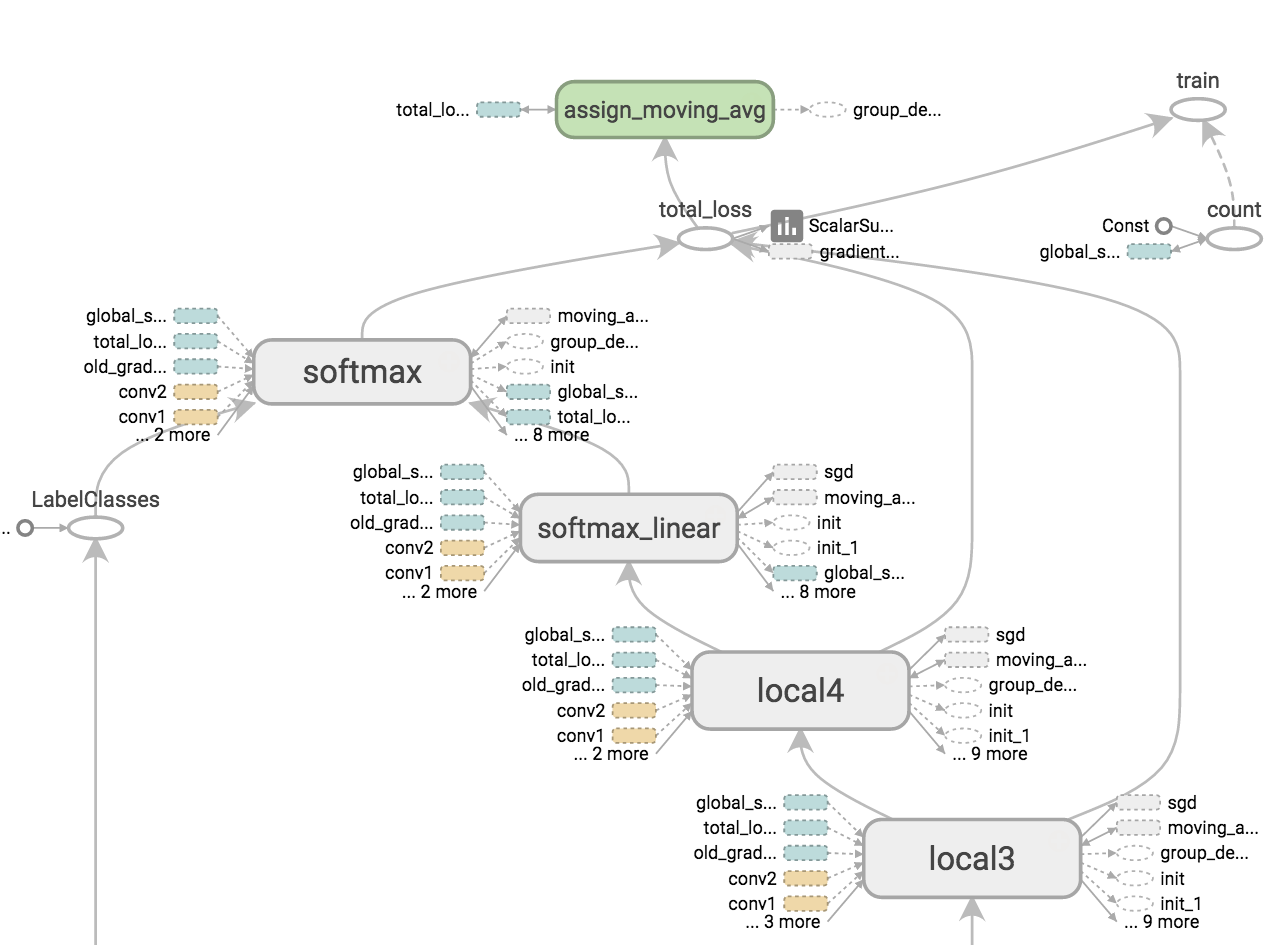}}
\caption{TensorBoard graph visualization of a convolutional neural network model}
\label{fig:graph-visualization}
\end{figure*}

The entire visualization is interactive: users can pan, zoom, and
expand grouped nodes to drill down for details. An example of the
visualization for the graph of a deep convolutional image model is
shown in Figure~\ref{fig:graph-visualization}.

\begin{figure*}
\centerline{\includegraphics[width=16cm]{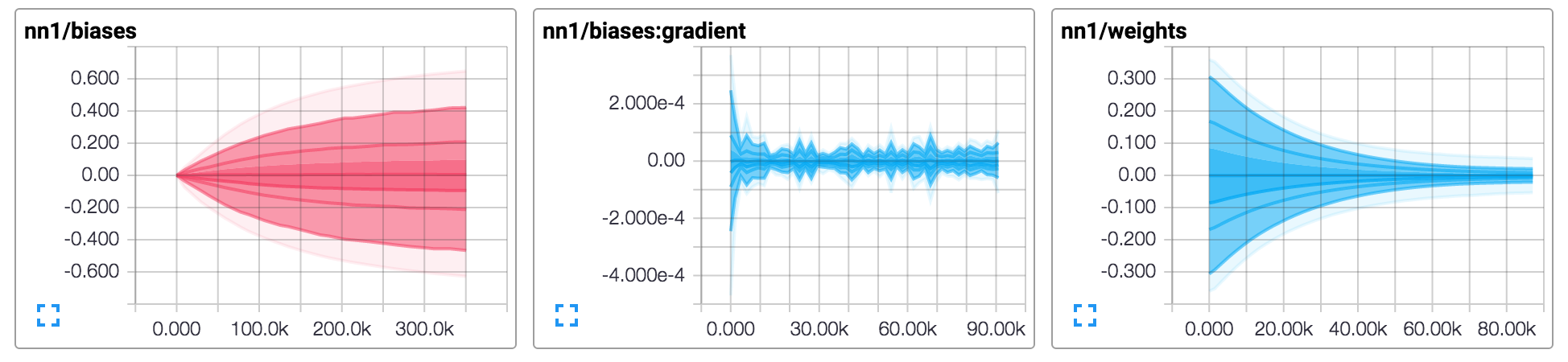}}
\caption{TensorBoard graphical display of model summary statistics time series data}
\label{fig:tensorboard-charts}
\end{figure*}

\subsubsection*{Visualization of Summary Data}

When training machine learning models, users often want to be able to
examine the state of various aspects of the model, and how this state
changes over time.  To this end, TensorFlow supports a collection of
different Summary operations that can be inserted into the graph,
including scalar summaries (e.g., for examining overall properties of
the model, such as the value of the loss function averaged across a
collection of examples, or the time taken to execute the computation
graph), histogram-based summaries (e.g., the distribution of weight
values in a neural network layer), or image-based summaries (e.g., a
visualization of the filter weights learned in a convolutional neural
network).  Typically computation graphs are set up so that Summary
nodes are included to monitor various interesting values, and every so
often during execution of the training graph, the set of summary nodes
are also executed, in addition to the normal set of nodes that are
executed, and the client driver program writes the summary data to a
log file associated with the model training.  The TensorBoard program
is then configured to watch this log file for new summary records, and
can display this summary information and how it changes over time
(with the ability to select the measurement of ``time'' to be relative
wall time since the beginning of the execution of the TensorFlow
program, absolute time, or ``steps'', a numeric measure of the number of
graph executions that have occurred since the beginning of execution
of the TensorFlow program).  A screen shot of the visualization of
summary values in TensorBoard is shown in Figure~\ref{fig:tensorboard-charts}.

\subsection{Performance Tracing}

We also have an internal tool called EEG (not included in the initial
open source release in November, 2015) that we use to collect and
visualize very fine-grained information about the exact ordering and
performance characteristics of the execution of TensorFlow graphs.
This tool works in both our single machine and distributed
implementations, and is very useful for understanding the bottlenecks
in the computation and communication patterns of a TensorFlow program.

Traces are collected simultaneously on each machine in the system from
a variety of sources including Linux kernel \texttt{ftrace}, our own
lightweight thread tracing tools and the CUDA Profiling Tools
Interface (CUPTI).  With these logs we can reconstruct the execution
of a distributed training step with microsecond-level details of every
thread-switch, CUDA kernel launch and DMA operation.

Traces are combined in a visualization server which is designed to
rapidly extract events in a specified timerange and summarize at
appropriate detail level for the user-interface resolution.  Any
significant delays due to communication, synchronization or
DMA-related stalls are identified and highlighted using arrows in the
visualization.  Initially the UI provides an overview of the entire
trace, with only the most significant performance artifacts
highlighted. As the user progressively zooms in, increasingly fine
resolution details are rendered.

\begin{figure*}
\centerline{\includegraphics[width=\textwidth]{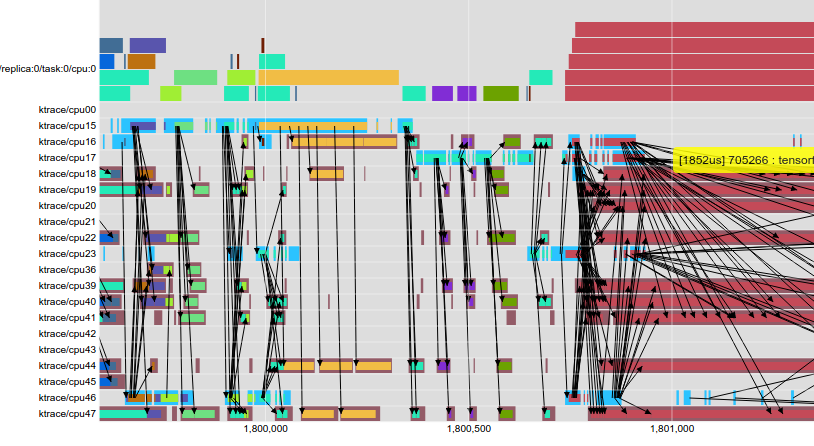}}
\caption{EEG visualization of multi-threaded CPU operations (x-axis is
  time in $\mu$s).}
\label{fig:eeg}
\end{figure*}
Figure~\ref{fig:eeg} shows an example EEG visualization of a model
being trained on a multi-core CPU platform. The top third of the
screenshot shows TensorFlow operations being dispatched in parallel,
according to the dataflow constraints. The bottom section of the trace
shows how most operations are decomposed into multiple work-items
which are executed concurrently in a thread pool. The diagonal arrows
on the right hand size show where queueing delay is building up in the
thread pool.
\begin{figure*}
  \centerline{\includegraphics[width=16cm]{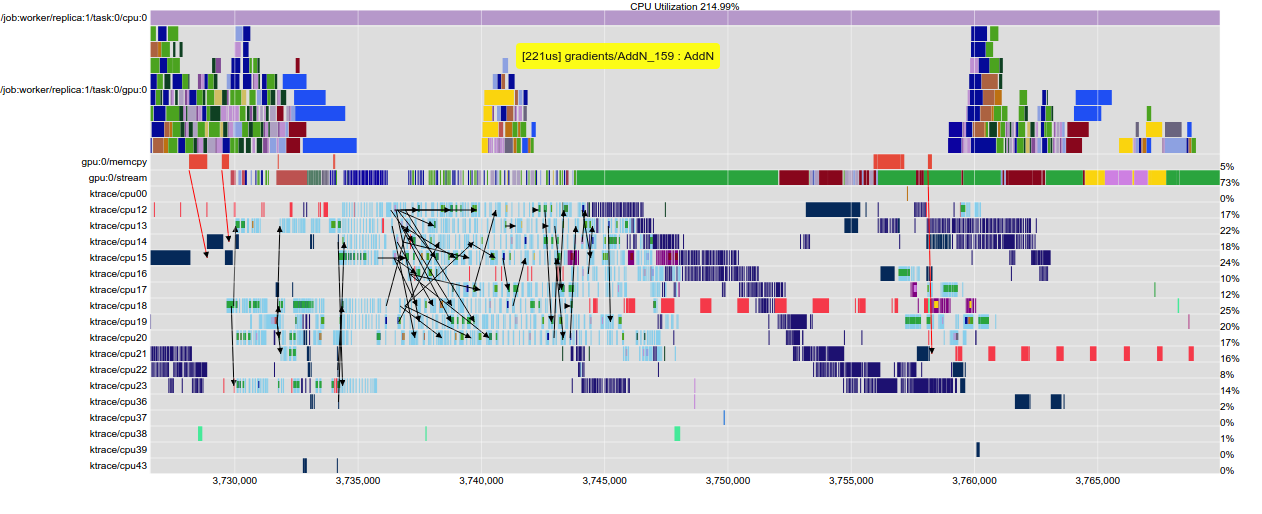}}
\caption{EEG visualization of Inception training showing CPU and GPU activity.}
\label{fig:eeg-imagenet-gpu}
\end{figure*}
Figure~\ref{fig:eeg-imagenet-gpu} shows another EEG visualization with
computation mainly happening on the GPU. Host threads can be seen
enqueuing TensorFlow GPU operations as they become runnable (the light
blue thread pool), and background housekeeping threads can be seen in
other colors being migrated across processor cores.  Once again,
arrows show where threads are stalled on GPU to CPU transfers, or
where ops experience significant queueing delay.

\begin{figure*}
  \centerline{\includegraphics[width=16cm]{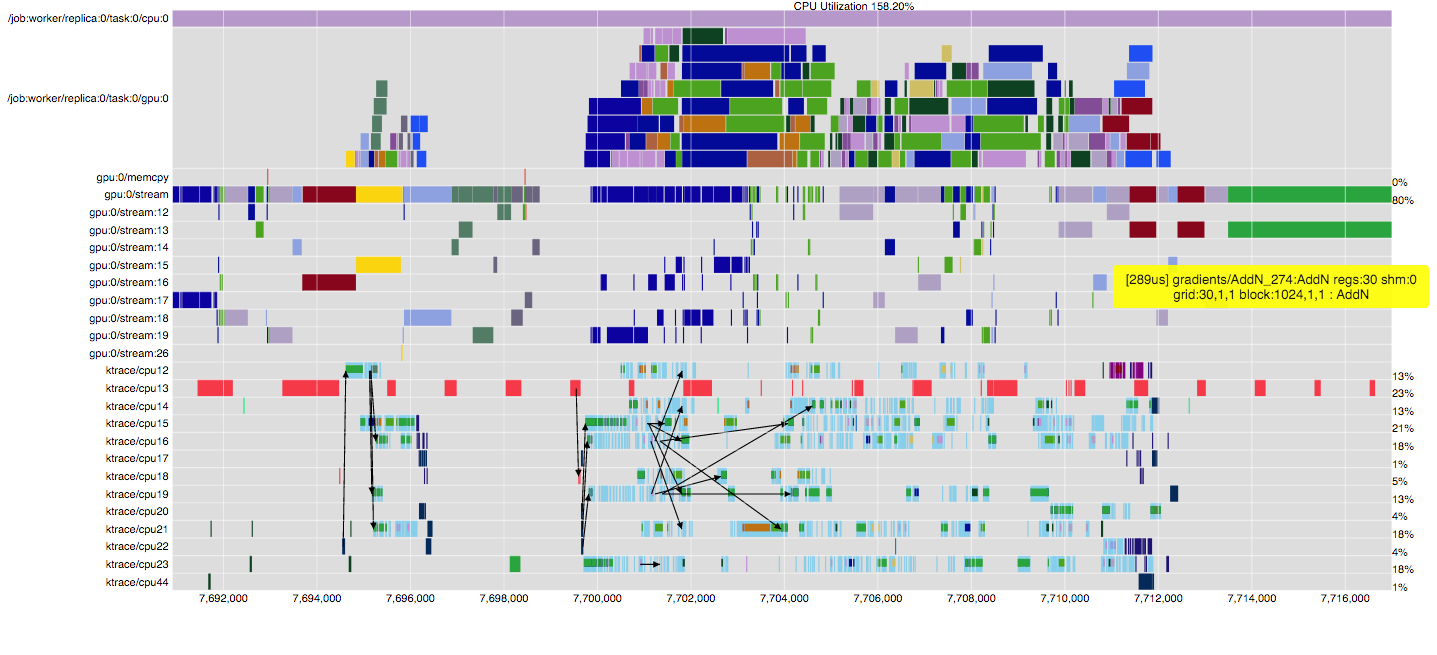}}
\caption{Timeline of multi-stream GPU execution.}
\label{fig:timeline-imagenet-8stream}
\end{figure*}
Finally, Figure~\ref{fig:timeline-imagenet-8stream} shows a more
detailed view which allows us to examine how Tensorflow GPU operators
are assigned to multiple GPU streams.  Whenever the dataflow graph
allows parallel execution or data transfer we endeavour to expose the
ordering constraints to the GPU device using streams and stream
dependency primitives.

\section{Future Work}
\label{sec:futurework}

We have several different directions for future work.  We will
continue to use TensorFlow to develop new and interesting machine
learning models for artificial intelligence, and in the course of
doing this, we may discover ways in which we will need to extend
the basic TensorFlow system.  The open source community may also come
up with new and interesting directions for the TensorFlow
implementation.

One extension to the basic programming model that we are considering
is a function mechanism, whereby a user can specify an entire subgraph
of a TensorFlow computation to be a reusable component.  In the
implementation we have designed, these functions can become reusable
components even across different front-end languages for TensorFlow,
so that a user could define a function using the Python front end, but
then use that function as a basic building block from within the C++
front-end.  We are hopeful that this cross-language reusability will
bootstrap a vibrant community of machine learning researchers
publishing not just whole examples of their research, but also small
reusable components from their work that can be reused in other contexts.

We also have a number of concrete directions to improve the
performance of TensorFlow.  One such direction is our initial work on
a just-in-time compiler that can take a subgraph of a TensorFlow
execution, perhaps with some runtime profiling information about the
typical sizes and shapes of tensors, and can generate an optimized
routine for this subgraph.  This compiler will understand the
semantics of perform a number of optimizations such as loop fusion,
blocking and tiling for locality, specialization for particular shapes
and sizes, etc.

We also imagine that a significant area for future work will be in
improving the placement and node scheduling algorithms used to decide
where different nodes will execute, and when they should start
executing.  We have currently implemented a number of heuristics in
these subsystems, and we'd like to have the system instead learn to
make good placement decisions (perhaps using a deep neural network,
combined with a reinforcement learning objective function).

\section{Related Work}
\label{sec:relatedwork}

There are many other systems that are comparable in various ways with
TensorFlow.  Theano~\cite{bergstra2010theano},
Torch~\cite{collobert2002torch}, Caffe~\cite{jia2014caffe},
Chainer~\cite{chainerorg} and the
Computational Network Toolkit~\cite{yu2014introduction} are a few systems
designed primarily for the training of neural networks.  Each of these
systems maps the computation onto a single machine, unlike the
distributed TensorFlow implementation.  Like Theano and Chainer, TensorFlow
supports symbolic differentiation, thus making it easier to define and work
with gradient-based optimization algorithms.  Like Caffe, TensorFlow
has a core written in C++, simplifying the deployment of trained models in
a wide variety of production settings, including memory- and
computation-constrained environments such as mobile devices.

The TensorFlow system shares some design characteristics with its
predecessor system, DistBelief~\cite{Dean-et-al-NIPS2012}, and with
later systems with similar designs like Project
Adam~\cite{chilimbi2014project} and the Parameter Server
project~\cite{parameterserverorg}.  Like DistBelief and Project Adam,
TensorFlow allows computations to be spread out across many
computational devices across many machines, and allows users to
specify machine learning models using relatively high-level
descriptions.  Unlike DistBelief and Project Adam, though, the
general-purpose dataflow graph model in TensorFlow is more flexible
and more amenable to expressing a wider variety of machine learning
models and optimization algorithms.  It also permits a significant
simplification by allowing the expression of stateful parameter nodes
as variables, and variable update operations that are just additional
nodes in the graph; in contrast, DistBelief, Project Adam and the
Parameter Server systems all have whole separate parameter server
subsystems devoted to communicating and updating parameter values.

The Halide system~\cite{ragan2013halide} for expressing image
processing pipelines uses a similar intermediate representation to the
TensorFlow dataflow graph.  Unlike TensorFlow, though, the Halide
system actually has higher-level knowledge of the semantics of its
operations and uses this knowledge to generate highly optimized pieces of code
that combine multiple operations, taking into account
parallelism and locality.  Halide runs the resulting computations only
on a single machine, and not in a distributed setting.  In future work
we are hoping to extend TensorFlow with a similar cross-operation
dynamic compilation framework.

Like TensorFlow, several other distributed systems have been developed
for executing dataflow graphs across a cluster.
Dryad~\cite{isard2007dryad} and Flume~\cite{chambers2010flumejava}
demonstrate how a complex workflow can be represented as a dataflow
graph. CIEL~\cite{murray2011ciel} and Naiad~\cite{murray2013naiad}
introduce generic support for data-dependent control flow: CIEL
represents iteration as a DAG that dynamically unfolds, whereas Naiad
uses a static graph with cycles to support lower-latency
iteration. Spark~\cite{zaharia2012resilient} is optimized for
computations that access the same data repeatedly, using ``resilient
distributed datasets'' (RDDs), which are soft-state cached outputs of
earlier computations. Dandelion~\cite{rossbach2013dandelion} executes
dataflow graphs across a cluster of heterogeneous devices, including
GPUs. TensorFlow uses a hybrid dataflow model that borrows elements
from each of these systems. Its dataflow scheduler, which is the
component that chooses the next node to execute, uses the same basic
algorithm as Dryad, Flume, CIEL, and Spark. Its distributed
architecture is closest to Naiad, in that the system uses a single,
optimized dataflow graph to represent the entire computation, and
caches information about that graph on each device to minimize
coordination overhead. Like Spark and Naiad, TensorFlow works best
when there is sufficient RAM in the cluster to hold the working set of
the computation. Iteration in TensorFlow uses a hybrid approach:
multiple replicas of the same dataflow graph may be executing at once,
while sharing the same set of variables. Replicas can share data
asynchronously through the variables, or use synchronization
mechanisms in the graph, such as queues, to operate
synchronously. TensorFlow also supports iteration within a graph,
which is a hybrid of CIEL and Naiad: for simplicity, each node fires
only when all of its inputs are ready (like CIEL); but for efficiency
the graph is represented as a static, cyclic dataflow (like Naiad).

\section{Conclusions}
\label{sec:conclusions}

We have described TensorFlow, a flexible data flow-based programming
model, as well as single machine and distributed implementations of
this programming model.  The system is borne from real-world
experience in conducting research and deploying more than one hundred
machine learning projects throughout a wide range of Google products
and services.  We have open sourced a version of TensorFlow, and hope
that a vibrant shared community develops around the use of TensorFlow.
We are excited to see how others outside of Google make use of
TensorFlow in their own work.

\section*{Acknowledgements}

The development of TensorFlow has benefitted enormously from the large
and broad machine learning community at Google, and in particular from
the suggestions and contributions from rest of the
Google Brain team and also from the hundreds of
DistBelief and TensorFlow users within Google.  Without a doubt, the
usability and functionality of TensorFlow has been greatly expanded by
listening to their feedback.

Many individuals have contributed to TensorFlow and to its open source
release, including
John~Giannandrea (for creating a supportive research environment),
Irina~Kofman and Phing~Turner (project management),
Bill~Gruber and David~Westbrook (technical writing),
Dave~Andersen, Anelia~Angelova, Yaroslav~Bulatov, Jianmin Chen, Jerjou~Cheng,
George~Dahl, Andrew~Dai,
Lucy~Gao,
mig~Gerard,
Stephan~Gouws, Naveen~Kumar, Geoffrey~Hinton, Mrinal~Kalarishnan, Anjuli~Kannan, Yutaka~Leon-Suematsu,
Frank~Li, Peter~Liu, Xiaobing~Liu, Nishant~Patil, Pierre~Sermanet, Noam~Shazeer, Jascha~Sohl-dickstein,
Philip~Tucker, Yonghui~Wu, Ke~Yang, and Cliff~Young (general contributions),
Doug~Fritz, Patrick~Hurst, Dilip~Krishnan, Daniel~Smilkov, James~Wexler, Jimbo~Wilson, Kanit~Ham~Wongsuphasawat, Cassandra~Xia, and the Big~Picture team (graph visualization),
Chris~Leary, Robert~Springer and the Stream~Executor team,
Kayur~Patel, Michael~Piatek, and the coLab team,
and the many others who have contributed to
the TensorFlow design and code base.

{\small
\bibliographystyle{plain}
\bibliography{tensorflow}
}

\end{document}